\documentclass[aps,prl,twocolumn,groupedaddress,amssymb]{revtex4}
\usepackage{graphicx}
\usepackage[figuresright]{rotating}
\begin{document}

\def\mdr{\mathit{\delta r}}

\title{On the Aizenman exponent in critical percolation}
\author{Lev N. Shchur \dag\ and Timofey Rostunov}
\affiliation{Landau Institute for Theoretical Physics, 142432 Chernogolovka,
Russia \\
\dag\ \sl e-mail: lev@itp.ac.ru}

\begin{abstract}
The probabilities of clusters spanning a hypercube of dimensions two to
seven along one axis of a percolation system under criticality were
investigated numerically. We used a modified Hoshen--Kopelman algorithm
combined with Grassberger's ``go with the winner" strategy for the site
percolation. We carried out a finite-size analysis of the data and found
that the probabilities confirm the Aizenman's proposal of the multiplicity
exponent for dimensions three to five. A crossover to the mean-field
behavior around the upper critical dimension is also discussed.
\end{abstract}


\maketitle

Percolation occurs in many natural processes, from electrical conduction
in disordered matter to oil extraction from field. In the latter, the
coefficient of oil extraction from oil sands (the ratio of the actually
extracted to the estimated oil) can be as much as 0.7 for light oil and as
low as 0.05 for viscous heavy oil. An increase in this coefficient by any
new point requires appreciable investment. Additional knowledge about the
percolation model could reduce the amount of additional investment.

A remarkable breakthrough in the theory of critical percolation was
established in the last decade thanks to a combination of mathematical
proofs, exact solutions, and large-scale numerical simulations. Recently,
Aizenman has proposed a new exponent that describs the probability
$P(k,r)$ of a critical percolation $d$-dimensional system with the aspect
ratio $r$ being spanned by at least $k$ clusters~\cite{Aiz97},
\begin{equation}
\ln P(k,r) \propto -\alpha_d\; k^\zeta\; r,
\label{Aiz-proposal}
\end{equation}
where $\alpha_d$ is a universal coefficient depending only on the
universality class, and $\zeta=d/(d-1)$.

In two dimensions, Aizenman's proposal~(\ref{Aiz-proposal}) was proved
mathematically~\cite{Aiz97}, confirmed numerically~\cite{SK1}, and derived
exactly~\cite{Cardy98} using conformal field theory and Coulomb gas
arguments. This exponent seems to be related to the exponents of
two-dimensional copolymers~\cite{Dup}. In three dimensions,
proposal~(\ref{Aiz-proposal}) was checked numerically in~\cite{Lev99} and,
more recently and more precisely, in~\cite{Grass02}.

\begin{figure}
\centering
\includegraphics[angle=0,width=\columnwidth,keepaspectratio]{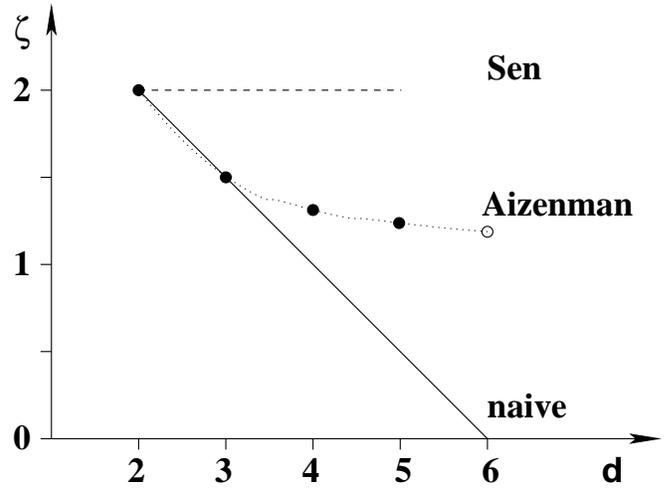}

\caption{Variation of Aizenman exponent $\zeta$ with the space dimension
$d$ as predicted by Aizenman (circles and dotted line), claimed by Sen
(dashed line) and discussed in the text (solid line).}

\label{fig.naive}
\end{figure}

The upper critical dimension of percolation is $d_c=6$, which follows from
the comparison of the exponents derived on the Cayley tree with those
satisfying scaling laws (see, e.g.,~\cite{StAh} and~\cite{BH}). The
fractal dimension $D_f$ of percolating critical clusters is equal to 4
above $d_c$, and the number of percolating clusters becomes infinite for
$d>d_c$. This fact would imply that $\zeta=0$ at $d=6$ if we supposed
(rather naively) that Aizenman's formula applies at the upper critical
dimension. Supposing that this is true and taking into account that the
values of $\zeta$ for $d=2$ and $d=3$ are, respectively, $2$ and $1.5$, we
can place all three points on the straight line $\zeta=(6-d)/2$, as
depicted in Fig.~\ref{fig.naive}. We can then estimate the respective
values of $\zeta$ for $d=4$ and $d=5$ to be $\zeta=1$ and $\zeta=0.5$;
these values are far from those predicted by Aizenman's formula giving
$4/3$ and $5/4$, respectively. In contrast, based on simulations,
Sen~\cite{Sen97} claims that $\zeta=2$ for all dimensions from two to
five.

The main purpose of our simulations is to estimate the exponents for the
dimensions from two to six with an accuracy sufficient for distinguishing
between the values predicted for $d=4$ and $d=5$ by the Aizenman's formula
and a naive application of cluster fractal-dimension arguments and by the
straight-line fit, as discussed above.

In the rest of the paper, we briefly summarize the highlights of our study,
then present some details of our research, and finally discuss the results
for the Aizenman exponent and the physics of a crossover from the Aizenman
picture to the mean-field picture.

Our main results can be summarized as follows:

1. {\em Modified combination of the Hoshen--Kopelman algorithm and
Grassberger's strategy.} We use the Hoshen--Kopelman (HK)
algorithm~\cite{HK} to generate clusters and Grassberger's ``go with the
winner" strategy~\cite{Grass02} to track spanning clusters. We add a new
tag array in the HK algorithm, which allows the reduction of the tag
memory order from $L^d$ to $L^{d-1}$, where $L$ is the linear size of the
hypercubic lattice. As a result, the amount of memory is about two orders
less for large values of $L$, and the program is about four times
faster---the complexity of the algorithm is compensated by the lower
memory capacity needed for for swapping to and from the auxiliary array.

2. {\em Efficient realization of combined shift-register random number
generators.} We use an exclusive-or ($\oplus$) combination $z_n$ of two shift
registers:
\begin{eqnarray}
x_n &=& x_{n-9689}\oplus x_{n-5502},
\nonumber
\\
y_n &=& y_{n-4423}\oplus y_{n-2325},\qquad z_n=x_n\oplus y_n
\label{eq.rng}
\end{eqnarray}
(see~\cite{RNG} and the references therein). We reduce the computational
time for generating random numbers by a factor 3.5 through an efficient
technical modification: we use the SSE command set that is available on
processors of the Intel and AMD series starting from the Intel Pentium III
and AMD Athlon XP.

3. {\em Extraction of the exponents for dimensions three to five.} We
first use finite-size analysis to estimate the logarithm of the
probability $P(k,r)$ in the limit of infinite lattice size $L$. We then
fit data as a function of the number of spanning clusters $k$ to obtain
the Aizenman exponent $\zeta$.

4. {\em Confirmation of Aizenman's proposal.} The estimates of the exponent
$\zeta$ for the dimensions $d=2,3,4$, and $5$ coincide well with those
proposed by Aizenman.

5. {\em Qualitative interpretation of Aizenman's conjecture.} Cardy
interpreted Aizenman's result qualitatively in two dimensions on the basis
of assumption that the main mechanism for reducing the number of
percolation clusters is that some of them terminate. The same result can
be derived for the cluster confluence (or merging) mechanism. This means
that, in low dimensions, the percolation clusters consist of a number of
closed paths (loops), while, in higher dimensions, clusters are more
similar to trees. Indeed, it is well known that the probability of
obtaining loop becomes lower for higher dimensions and goes to zero in
the limit of infinite dimensions (Cayley tree)~\cite{StAh,BH}.

6. {\em Crossover to mean-field behavior.} We found evidence that the
probability of clusters spanning a hypercubic lattice tends to unity in
the limit of high dimensions, as it follows from the well-accepted
picture. We did not find any dramatic changes in the probabilities around
the upper critical dimension $d_c=6$, but rather found evidence for a
crossover. Therefore, Aizenman's formula~(\ref{Aiz-proposal}) can also
apply to dimensions higher (but not too much higher) than the upper
critical dimension and describe approximately the probabilities of
spanning clusters in large, though finite-size systems.

We follow with the details of the critical percolation, simulations, and
data analysis.

\begin{figure}
\centering
\includegraphics[angle=270,width=\columnwidth,keepaspectratio]{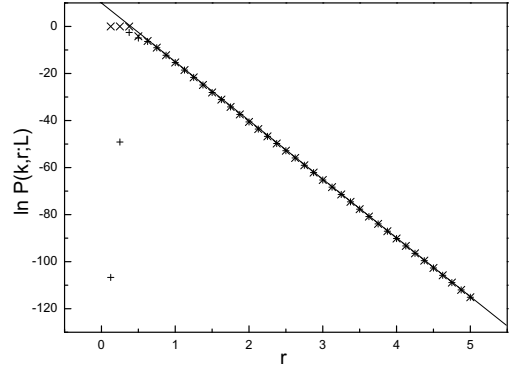}

\caption{The logarithms of the probabilities of exactly $k$ clusters
$P(k,r;L)$ ($+$) and of at least $k$ clusters $P_+(k,r;L)$ ($\times$) for the
dimension $d=4$ and the number of clusters $k=5$ as functions of the aspect
ratio $r$. The linear size of the hyperrectangle is $L=16$. The solid line is
the linear approximation to $\ln P(k,r;L)$ on the interval $r=[1.5;5.0]$.}

\label{fig.pr}
\end{figure}

{\sf Spanning probability.}

\noindent We can define the probability $P(k,r;L)$ that $k$ clusters
traverse a $d$-dimensional hyperrectangle $[0,L]^{d-1}\times[0,Lr]$ in the
$Lr$ direction~\cite{Aiz97}. Provided that the scaling limit exists (this
was proved recently by Smirnov for the percolation in
plane~\cite{Smirnov}), the probability $P(k,r)$ can be defined as the
limit of $P(k,r;L)$ as $L\to\infty$. Aizenman proposed that $P(k,r)$
should behave according to~(\ref{Aiz-proposal}) in dimensions from three
to five. The validity of formula~(\ref{Aiz-proposal}) for the percolation
in plane was well established in \cite{Aiz97,Cardy98,SK1}.

Numerical results (\cite{Lev99} and~\cite{Grass02}) for the exponent $\zeta$
for critical percolation on cubic lattices seems to confirm Aizenman's
proposal for the value of $\zeta=1.5$

Actually, we could consider the probability $P(k,r)$ as the probability of
obtaining $k$ clusters at the distance $r$ from the left side of the
hyperrectangle if clusters grow to the right. Only two processes can
change the number of clusters: cluster merging and cluster terminating.

The differential $dP$ of the probability is
\begin{equation}
dP\propto P(k,r)\;k^{1/(d-1)}\;k\;dr,
\label{difP}
\end{equation}
where the right-hand side represents the product of the probability
$P(k,r)$ and the differential of the total border hyperarea of $k$
clusters, each with the hyperarea differential $k^{1/(d-1)}\;dr$. This
expression follows from the fact that area unit of measure is proportional
to the characteristic transverse length of ``infinite" clusters.
Therefore, the transverse area remains constant as $k$ changes, while the
longitudinal length increment in these units is $\propto k^{1/(d-1)}dr$.
Integrating~(\ref{difP}), we recover probability~(\ref{Aiz-proposal}).
Thus, $P(k,r)$ describes the probability that $k$ clusters do not merge
together.

The same probability could be obtained by the process of cluster
termination, as given by Cardy in plane~\cite{Cardy98}, which can easily
be extended to dimensions $d>2$.

This means that the exponent $\zeta$ cannot be larger than the one
proposed by Aizenman, and $\zeta=d/(d-1)$ is the upper bound for the
exponent.

\begin{table}
\begin{tabular}{cccccll}
\hline
d & k & $L_{min}$ & $L_{max}$ & $\delta L$& $ \qquad p_c$ & Ref.\\ \hline
2 & 1-5 & 16 & 256 &16-32 & 0.59274621(13) & \cite{pc2} \\ 
3 & 1-6 & 8 & 64 & 4,8 & 0.3116080(4) & \cite{pc3} \\ 
4 & 1-6 & 8 & 48-56 & 4,8 & 0.196889(3) & \cite{pc45}\\ 
5 & 1-6 & 4 & 32,24 & 4,8 & 0.14081(1) & \cite{pc45}\\ 
6 & 1-6 & 4 & 15-16 & 3-5 & 0.109017(2) & \cite{pc6}\\ 
7 & 1-4 & 4 & 10 & 1 & 0.0889511(9) & \cite{pc6} 
\end{tabular}

\caption{Minimal $L_{min}$ and maximal $L_{max}$ linear sizes of the
percolation lattice and the interval $\delta L$ between two consecutive
values of $L$ depending on the dimension $d$ and number of clusters $k$.
The values of $p_c$ are taken from the references in the last column.}
\label{tab.L}
\end{table}

{\sf Algorithms and realizations.}

\noindent The classical realization of the HK algorithm~\cite{HK} requires
memory for two major structures: an array for keeping a
$(d{-}1)$-dimensional cluster slice and a tag array. The total memory
required by the algorithm is $\propto L^{d-1}+p_crL^d$, where $p_c$ is the
site percolation threshold value. Therefore, for large $rL$, one of the
main advantages of the HK algorithm (i.e., relatively low memory
consumption) is negated by the second term. Our modification of the
original algorithm allows the memory for the tag array to be reduced to
about $3p_cL^{d-1}$.

Instead of keeping all tags in memory and selecting new tags with
increasing tag numbers, we create two arrays, of which one keeps the tag
value and the other one keeps the number $N$ of the slice where the
corresponding tag was last used. When we build a cluster, we update this
array with $N=N_\mathrm{current}$ for the tags used. If
$N<N_\mathrm{current}-1$, then this tag is not on the front surface of the
sample, and it will never be used again so that we can, therefore, reuse
it. We note that cluster size information should be taken into account
before reusing the associated tag, if the size information is required.

We use the ``go with the winner" strategy~\cite{Grass02} as follows. If
the system has $k$ spanning clusters for some aspect ratio
$r=n\mathit{\delta r}$, it is stored in memory and is grown for $\mdr$. If
the resulting configuration has $k$ spanning clusters, it is stored, and
the growth process continues. Otherwise, we return to the previously saved
state. Using this procedure, we calculate the probability $P_i(\mdr)$ that
the system propagates at the distance $\mdr$ from the position
$r=(i-1)\mdr$. Finally, we obtain $P(r=n\mdr)=\prod_{i=1}^nP_i(\mdr)$. By
choosing sufficiently small values of $\mdr$, we can achieve rather high
probabilities of $P_i(\mdr)$ (which can be determined from a few
realizations), while the total probability may be very small (down to
$\propto 10^{-100}$ in our case).

The random number generator was optimized for the SSE instruction set as
follows. Because the length of all four RNG legs is $\{a|b\}_{\{x|y\}}>4$,
the $n$th step of the RNG does not intersect with the ($n{+}3$)th step.
Therefore, we can pack four consecutive 32-bit values of
\{$x_{n-\{a_x|b_x\}}$\} and \{$y_{n-\{a_y|b_y\}}$\} into 128-bit XMM
registers, process them simultaneously (see Eq.~(\ref{eq.rng})), and thus
obtain $z_n$, $z_{n+1}$, $z_{n+2}$, and $z_{n+3}$ within one RNG cycle.

\begin{figure}
\includegraphics[angle=270,width=\columnwidth,keepaspectratio]{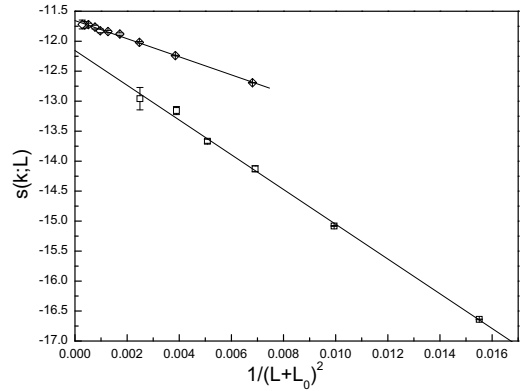}
\caption{Plot of $s(k;l)$ for $k=3$ clusters in the dimension four as
a function of $1/(L+L_0)$ with the fitting parameter $L_0=4.12$ ($\square$)
and for $k=4$ clusters in the dimension six with $L_0=4.03$ ($\diamond$).
Straight lines result from the fitting to the corresponding data as discussed
in the text.}
\label{fig.ldep}
\end{figure}

{\sf Data analysis.}

\noindent The lattice size was varied from $L_{min}$ to $L_{max}$ with the
step $\delta L$. In Table~\ref{tab.L}, particular values of the simulation
parameters are presented together with the interval of the number of
clusters $k$ depending on the dimension $d$. The direct result of the
simulations is the probabilities $P(k,r;L)$ that exactly $k$ clusters
connect two opposite surfaces (separated by the distance $rL$) of the
rectangle with size $L^{d-1}$ in the ``perpendicular" direction in which
we apply periodic boundary conditions. We use the values of the site
percolation thresholds on hypercubic lattices from~\cite{pc2}--\cite{pc6},
as shown in Table~\ref{tab.L}.

\begin{table}
\begin{tabular}{clllll}
\hline
 &\multicolumn{5}{c}{d} \\ \cline{2-6}
$k$ & \quad3 & \quad4 & \quad5 & \quad6 & \quad7 \\ \hline
1 & -1.377(1) & -1.774(3) & -1.859(9) & -1.76(2) & -1.48(4) \\ 
2 & -6.919(6) & -6.330(15) & -5.57(6) & -4.73(8)& -3.55(11) \\
3 & -13.655(15) & -11.64(4) & -9.95(12) & -8.27(12) & -6.25(16) \\ 
4 & -21.47(3) & -17.77(6) & -14.65(20)& -11.95(25)& -9.3(3) \\ 
5 & -30.23(3) & -24.02(8) & -19.9(3) & -15.75(30)& \\ 
6 & -40.02(6) & -31.0(1) & -25.0(3) & -22.7(2) & \\ 
\end{tabular}
\caption{Values of $s(k)$ for different numbers of clusters $k$
and dimensions $d$ for site percolation on hypercubic lattices with periodic
boundary conditions in directions perpendicular to the spanning direction.}
\label{alpha}
\end{table}

Data analysis consists of three steps. First, we compute the slope $s(L)$
of $\ln P(k,r;L)$ for a given dimension $d$, number of clusters $k$, and
linear lattice size $L$. An example of such a function is given in
Fig.~\ref{fig.pr} for $\ln P(5,r;16)$ in the dimension four. We also plot
the logarithm of the probability $P_+(k,r;L)=\sum_{k'\geqslant
k}P(k',r;L)$ of the event that, {\em at least}, $k$ clusters span the
(hyper)rectangle at the distance $rL$. To calculate $s(L)$, we use data
only in the interval of the aspect ratio $r$ between $1.5$ and $5$. We
note that the probability of five clusters spanning a rectangle with the
linear size $L=16$ at the distance $5\cdot 16=80$ is extremely small
$\approx 10^{-52}$.

Second, we compute probabilities in the limit of an infinite system size $L$,
fitting slopes $s(k)$ with the expression (see Fig.~\ref{fig.ldep})
\begin{equation}
s(k;L)=s(k)+\frac{B}{(L+L_0)^t},
\label{fit-L}
\end{equation}
where $B$, $t$, and $L_0$ are fitting
parameters~\cite{Ziff92,Lev99,Ziff02}. The resulting values of the slopes
$s(k)$ are presented in Table~\ref{alpha}. The number of runs used to
compute each particular entry in Table~\ref{alpha} varied from $10^6$
to several tens for higher dimensions.

\begin{table}
\begin{tabular}{llll}
\hline
$k$ & this paper & exact from \cite{Cardy98} &
 from \cite{Lev99} \\ \hline
 1 & -0.6541(5) & -0.6544985 & -0.65448(5) \\
 2 & -7.855(3) & -7.85390 & -7.852(1) \\ 
 3 & -18.32(1) & -18.3260 & -18.11(15) \\ 
 4 & -32.99(3) & -32.9867 & \\ 
 5 & -51.83(2) & -51.8363 & \\
\end{tabular}
\caption{Values of $s(k)$ in two dimensions for different $k$
calculated in this paper, using the exact Cardy formula~\cite{Cardy98}, and
estimated in~\cite{Lev99} for site percolation on a tube.}
\label{tab.d2}
\end{table}

We checked the accuracy of our simulations, as well as the validity of the
approach in general for site percolation on a square lattice.
Table~\ref{tab.d2} shows a comparison of our results for the slope $s$
with the exact values and with earlier simulations, in which the other
modification of the HK algorithm, but not the Grassberger strategy, was
used. We note that our results coincide well with the exact results and
give a higher accuracy for larger values of $k$ in comparison with the
previous numerical results, despite the smaller computation time used. Our
data for $k=1$ is less accurate because of the smaller statistics ($10^6$
runs, compared to $10^8$ samples in~\cite{Lev99}). This is a direct
demonstration of the effectiveness of the Grassberger strategy for large
values of $k$.

Finally, we use values in Table~\ref{alpha} to determine the Aizenman
exponent $\zeta$ by fitting data in each column to
\begin{equation}
s=A(k^2-k_0)^{p/2}
\label{eq-fit1}
\end{equation}
in two and three dimensions, as proposed by Grassberger~\cite{Grass02},
and to
\begin{equation}
s=A(k^p-k_0)
\label{eq-fit2}
\end{equation}
in higher dimensions. Here, $A$, $k_0$, and $p$ are fitting parameters. We
take only the leading behavior in $k$ into account.

\begin{table}
\begin{tabular}{llll}
\hline
$d$ & $\quad A $ & $\quad k_0$ & $\quad p$ \\ \hline \hline
2 & 2.090(4) & 0.244(5) & 2.0012(10) \\ 
 & 2.0940(5) & 0.2489(7) & 2 \\ \hline 
3 & 2.81(4) & 0.64(4) & 1.489(7) \\ 
 & 2.757(2) & 0.587(3) & 3/2 \\  \hline
4 & 3.06(20) & 0.41(6) & 1.315(30) \\ 
 & 2.949(5) & 0.373(3) & 4/3 \\ \hline 
5 & 2.8(1) & 0.40(4) & 1.24(3) \\ 
 & 2.78(2) & 0.38(2) & 5/4 \\ \hline 
6 & 2.8(8) &0.5(3) & 1.12(14) \\ 
 & 2.41(5) &0.33(6) &6/5 \\ \hline
7 & 1.4(10) & 0.08(116) & 1.4(4)\\ 
 &2.03(12)&0.50(13)&7/6
\end{tabular}
\caption{Values of the fitting parameters $A$ and $k_0$ and the power $p$ as
defined in Eqs.~(\ref{eq-fit1}) and~(\ref{eq-fit2}) for the dimension $d$.}
\label{tab.res}
\end{table}

{\sf Spanning, proliferation, and crossover to mean-field behavior.}

\noindent The results of the final fit to (\ref{eq-fit1}) and
(\ref{eq-fit2}) are shown in Table~\ref{tab.res}. The second row for each
particular dimension $d$ is the fit with the power $p$ fixed to the
Aizenman exponent value. This is done to check the fit stability. Indeed,
the values of $A$ and $k_0$ coincide within one standard deviation for the
dimensions two to five.

\begin{figure}
\includegraphics[angle=270,width=\columnwidth,keepaspectratio]{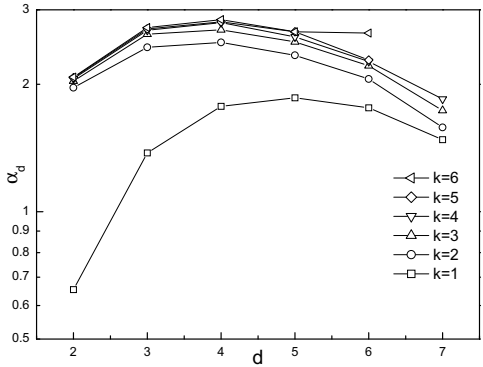}
\caption{The coefficient $\alpha_d$ (as a function of the dimension $d$)
extracted from the probabilities $P(k,r)$ for different numbers of clusters
$k$.}
\label{fig.coeff2}
\end{figure}

The larger deviations of parameters for the dimensions six and seven may
be attributed to appearance of cluster proliferation---the number of
clusters is known~\cite{Aiz97} to grow as $L^{d-6}$ in dimensions
$d>d_c=6$. We plot the coefficient $\alpha_d$ (defined by
expression~(\ref{Aiz-proposal})) in Fig.~\ref{fig.coeff2} as a function of
the dimension $d$. The probability of exactly one cluster spanning at the
given distance $r$ becomes smaller as the dimension increases from two to
five and larger for larger dimensions, as can be seen from the first row
($k=1$) of Table~\ref{alpha} and from the lower curve dependence in
Fig.~\ref{fig.coeff2}.  For any fixed $d$, the value of $\alpha_d$
approaches some limit for the dimensions two to five and $k>2$, which
suggests the value of the corrections to the leading behavior in $k$ (see
Eqs.~(\ref{eq-fit1}) and (\ref{eq-fit2})).

The fact that the value of $\zeta$, which we formally extracted from our
data for $d=6$, more or less coincides with $\zeta=d/(d-1)=6/5$, as
formally calculated using the Aizenman expression, may be interpreted as
an indication that the number of clusters depends logarithmically on the
lattice size $L$. One can expect that the logarithmic behavior is visible
only for somewhat larger values of $L$ than we have used so far (see
Table~\ref{tab.L}). With the values of $L$ of the order we have used in
simulations, we see effectively the same picture as for the lower
dimensions---clusters spann according to the Aizenman formula. This means
that at small (or moderate) values of $L$, the main mechanism is as
discussed above: cluster merging and terminating. And only at sufficiently
large system sizes we will see cluster proliferation. An indication of
that can be seen from the values of $\alpha_d$ in the dimension seven in
Fig.~\ref{fig.coeff2}. The probabilities become closer, and this can be
attributed to cluster proliferation and treated as a crossover to the
mean-field behavior.

{\sf Discussion.}

\noindent The results have shown the validity of the Aizenman's proposal
in the dimensions three to five (results on plane were already proved
rigorously) and do not support Parongama Sen claims based on their
simulations (Fig.~\ref{fig.naive}).  We have found evidence for cluster
proliferation for the dimension seven. The analysis can be extended to the
number of spanning clusters to distinguish exponential decay with the
system size of the number of clusters for the dimension five, logarithmic
growth of them for the dimension six, and linear growth for the dimension
seven. The same technique can be used to establish numerically such a
crossover to the mean-field picture, although a significantly longer
computational time, than we used, is needed for this. In fact, the linear
growth of the multiplicity of spanning clusters for seven-dimensional
critical percolation was confirmed numerically in preprint~\cite{ACF}
posted at {\em arXiv} preprint library a few days after our
cond-mat/0207605.

\acknowledgments

We acknowledge useful discussions of algorithms with P.~Grassberger and
R.~Ziff. Our special thanks to S.~Korshunov and G.~Volovik for the
discussion of the results. This work supported by grants from 
Russian Foundation for Basic Research.

\end{document}